\documentclass[11pt]{article}
\usepackage{amssymb,amsmath,amsfonts,latexsym}
\usepackage[mathscr]{eucal}
\date{}
\title{{Energy-momentum distribution of a general plane symmetric spacetime in metric $f(\textmd{R})$ gravity}}
\author{\bf{Morteza Yavari}\\\
\small{Department of Physics, Islamic Azad University, Kashan Branch, Kashan, Iran}}
\begin{document}
\maketitle \noindent\small{In this paper, the exact vacuum solution of a general plane symmetric spacetime is investigated in metric $f(\textmd{R})$ gravity with the assumption of constant Ricci scalar. For this solution, we have studied the generalized Landau-Lifshitz energy-momentum complex in this theory to determine the energy distribution expressions for some specific $f(\textmd{R})$ models. Also, we show that these models satisfy the constant curvature condition.}\vspace{.15cm}\vspace{.15cm}
\large
\section{Introduction}
\noindent
The energy localization is still an unsolved problem in the framework of general relativity. A considerable amount of work has been devoted to study of the energy localization. For solving this problem, Einstein [1] introduced the energy-momentum pseudotensors. He formulated the energy-momentum conservation law as $\frac{\partial(\sqrt{-\textrm{g}}\,(\textrm{T}^\mu_\nu+t^\mu_\nu))}{\partial x^\mu}=0$, where $t^\mu_\nu$ is called the gravitational field pseudotensor. Many authors like Papapetrou [2], Bergmann [3], Goldberg [4] and Weinberg [5] have studied the energy-momentum complexes and covariant conservation laws. The most of these studies were restricted in Cartesian coordinates. M{\o}ller [6] was the first who describe the energy-momentum complexe in any coordinate system. Landau-Lifshitz [7] presented the energy-momentum complex in the geodesic coordinate system. Chang et al. [8] showed that any energy-momentum complex is connected with a Hamiltonian boundary term. This shows that we can consider the energy-momentum as quasi-local. The quasi-local energy were investigated by several authors (see e.g. [9,10]). Cooperstock and Sarracino [11] proved that the localization of energy in the spherical systems is the necessary condition for the localizable in any system. Aguirregabiria et al. [12] proved that the different energy-momentum complexes could give the same energy distribution for any Kerr-Schild spacetime. Recently, a number of authors have tried to solve the problem of energy localization via the modified theories of gravity.

Different data from the recent astrophysical observations such as Super-Nova Ia [13,14], Cosmic Microwave Background Radiations [15,16] and Wilkinson Microwave Anisotropy Probe [17] have indicated that the expansion of universe is currently accelerating. The standard general theory of relativity can not describe the accelerated expansion. Based on these data, physicists now believe that the most part of universe contains dark energy with negative pressure, in which this energy constrain the cosmic expansion [18-20]. One of the seriously approaches which may help to explain the origin of dark energy is to modify the general theory of relativity. The modified theories of gravity, such as $f(\textmd{R})$ gravity, have gained a lot of interest in recent years. In these theories, the geometrical part of Einstein-Hilbert action is modified by adding the higher-order curvature invariants [21]. Stelle [22] showed that the higher-order actions are renormalizable. Hence, modifying of Einstein-Hilbert action is a possible approach to make a renormalized theory of gravity [23]. Among the modified theories, the $f(\textmd{R})$ gravity seems to be an attractive model which is relatively simple but has many applications in gravity, cosmology and high energy physics [24]. In this theory, a general function of Ricci scalar as $f(\textmd{R})$ is replaced instead of R in the Einstein-Hilbert action, first discussed by Buchdahl [25]. Nojiri and Odintsov [26-28] showed that the modified theories of gravity provide a natural gravitational alternative way for dark energy. Nojiri and Odintsov [26-28] and Faraoni [29] have shown that the some $f(\textmd{R})$ theories can pass the Solar System tests.

Multam\"{a}ki et al. [30] studied the energy-momentum complexes in metric $f(\textmd{R})$ gravity. They generalized the energy-momentum complexe in metric $f(\textmd{R})$ gravity for constant curvature solutions. Sharif and Shamir [31] found the energy densities for some static plane symmetric solutions by using the generalized Landau-Lifshitz energy-momentum complex. In the present paper, we would like to extend this analysis for a general plane symmetric spacetime.

This paper is organized as follows: In section 2, the field equations in metric $f(\textmd{R})$ gravity are discussed. In section 3, the vacuum solutions of a general plane symmetric spacetime for constant curvature are found. In section 4, we firstly give a brief introduction about the generalized Landau-Lifshitz energy-momentum complex in the framework of $f(\textmd{R})$ gravity. Then, the energy distribution for the obtained solutions in section 3 are computed for a number of commonly considered $f(\textmd{R})$ theories. In the last section, we conclude the results.
\section{Field equations in $f(\textmd{R})$ gravity}
In this section, we give a brief review of the modified field equations in metric $f(\textmd{R})$ gravity. There are two formalisms which are applied to obtain the field equations in $f(\textmd{R})$ gravity. One is the metric formalism while the another approach is Palatini formalism. The modified field equations obtained by these two formalisms are not the same in general. The metric formation of this theory has been studied by a number of authors (see e.g. [27]). The metric and Palatini $f(\textmd{R})$ gravities have recently been reviewed in detail by Capozziello and Francaviglia [32], Sotiriou and Faraoni [33]. Olmo [34] has reviewed the recent literature on modified theories of gravity in Palatini approach.\\
The action for $f(\textmd{R})$ gravity coupled with matter is given by$\footnote{The gravitational units with c=G=1 are used.}$
\begin{eqnarray}
\textrm{S}=\frac{1}{16\pi}\int d^4x\sqrt{-\textrm{g}}\,f(\textmd{R})+\textrm{S}_m,
\end{eqnarray}
where $f(\textmd{R})$ is a general function of Ricci scalar and $\textrm{S}_m$ represents the action associated with the
matter fields. The field equations are reached by varying the above action with respect to the metric tensor $\textmd{g}_{\mu\nu}$,
then they are given by
\begin{eqnarray}
\textmd{R}_{\mu\nu}-\frac{1}{2}\,\textmd{g}_{\mu\nu}\textmd{R}=\textmd{T}^\textmd{g}_{\mu\nu}+8\pi\textmd{G}\frac{\textmd{T}^m_{\mu\nu}}{\,\textmd{F}(\textmd{R})},
\end{eqnarray}
in which $\textmd{T}^\textmd{g}_{\mu\nu}$ is the geometric
energy-momentum tensor and it defines as$\footnote{$\nabla_\mu$ is the
covariant derivative associated with the Levi-Civita connection
of the metric.}$
\begin{eqnarray}
\textmd{T}^\textmd{g}_{\mu\nu}=\frac{1}{\,\textmd{F}(\textmd{R})}\left\{\frac{1}{2}\,\textmd{g}_{\mu\nu}\left(f(\textmd{R})-\textmd{F}(\textmd{R})\textmd{R}\right)
+\nabla^\alpha\nabla^\beta\textmd{F}(\textmd{R})(\textmd{g}_{\alpha\mu}\textmd{g}_{\beta\nu}-\textmd{g}_{\mu\nu}\textmd{g}_{\alpha\beta})\right\},
\end{eqnarray}
with
$\textmd{F}(\textmd{R})\equiv\displaystyle\frac{\,df(\textmd{R})}{d\textmd{\textmd{R}}}$
and $\textmd{T}^m_{\mu\nu}$ is the standard matter stress-energy tensor derived from the matter action. For the vacuum solutions, the field equations become
\begin{eqnarray}
\textmd{F}(\textmd{R})\textmd{R}_{\mu\nu}-\frac{1}{2}\,f(\textmd{R})\textrm{g}_{\mu\nu}-\nabla_\mu\nabla_\nu
\textmd{F}(\textmd{R})+\textrm{g}_{\mu\nu}\Box\textmd{F}(\textmd{R})=0,
\end{eqnarray}
where $\Box\equiv\nabla^\mu\nabla_\mu$ is the d'Alembertian. Contracting the field equations, gives the following relation between
$f(\textmd{R})$ and its derivative
\begin{eqnarray}
\textmd{F(R)}\textmd{R}-2f(\textmd{R})+3\Box\textmd{F(R)}=0,
\end{eqnarray}
which will be used later to simplify the field equations and to determine the function of Ricci scalar. For constant curvature solutions ($\textmd{R}=\textmd{R}_0$), this equation reduces to
\begin{eqnarray}
\textmd{F}(\textmd{R}_0)\textmd{R}_0-2f(\textmd{R}_0)=0.
\end{eqnarray}
This condition is very important for checking the acceptability of $f(\textmd{R})$ models.
\section{The plane symmetric vacuum solutions}
The study of plane symmetric solutions in Einstein theory has a long history. The general vacuum solution of the plane symmetric model was first considered by Taub more than 60 years ago. A generalization of this spacetime with cosmological constant was first obtained by Novotn\'{y} and Horsk\'{y} [35]. In recent years, the plane symmetric spacetimes have been discussed extensively in general relativity by many authors. Sharif and Shamir [36] studied the constant curvature vacuum solutions of plane symmetric spacetime in metric $f(\textmd{R})$ gravity. Yavari [37] investigated a complete set of the exact vacuum solutions of the plane symmetric spacetime for two cases $\textmd{R}=\textmd{constant}$ and $\textmd{R}\not=\textmd{constant}$ in metric $f(\textmd{R})$ gravity. In this section, we find the exact solutions of vacuum field equations for a general plane symmetric spacetime in metric $f(\textmd{R})$ gravity. We consider the line element of plane symmetric spacetime in Cartesian coordinates given by
\begin{eqnarray}
ds^2=-adt^2+b(dx^2+dy^2)+cdz^2,
\end{eqnarray}
where $a$, $b$ and $c$ are unknown functions of $z$. The corresponding Ricci scalar is
\begin{eqnarray}
\textrm{R}=\frac{1}{2c}\left\{2\frac{\,a''}{a}+4\frac{\,b''}{b}-(\frac{\,a'}{a})^2-(\frac{\,b'}{b})^2+2\frac{\,a'}{a}\frac{\,b'}{b}-\frac{\,a'}{a}\frac{\,c'}{c}-2\frac{\,b'}{b}\frac{\,c'}{c}\right\},
\end{eqnarray}
here prime denotes derivative with respect to $z$. Next, by applying the equation (5), the vacuum field equations take the following form
\begin{eqnarray}
\nabla_\mu\nabla_\nu
\textmd{F}-\textmd{F}\textmd{R}_{\mu\nu}=\frac{\Box
\textmd{F}-\textmd{R}\textmd{F}}{4}\,\textrm{g}_{\mu\nu},
\end{eqnarray}
since the metric only depends on the coordinate $z$, this equation is a set of differential equations for functions
$a(z)$, $b(z)$ and $c(z)$. In this case both sides are diagonal and so, we have four equations. From the equation (9) it is clear that the combination
$\textmd{M}_\mu\equiv\frac{\,\nabla_\mu\nabla_\mu\textmd{F}-\textmd{F}\textmd{R}_{\mu\mu}}{\textmd{g}_{\mu\mu}}$ (with fixed indices) is
independent of the index $\mu$ and so, we have $\textmd{M}_\mu=\textmd{M}_\nu$ for all $\mu$ and $\nu$, [38]. From the last consequence, two following independent field equations are obtained
\begin{eqnarray}
&&2(\frac{\,a'}{a}-\frac{\,b'}{b})\textmd{F}'-\left\{2\frac{\,a''}{a}-2\frac{\,b''}{b}-(\frac{\,a'}{a})^2+\frac{\,a'}{a}\frac{\,b'}{b}-\frac{\,a'}{a}\frac{\,c'}{c}+\frac{\,b'}{b}\frac{\,c'}{c}\right\}\textmd{F}=0,\\
&&2\textmd{F}''-(\frac{\,a'}{a}+\frac{\,c'}{c})\textmd{F}'-\left\{2\frac{\,b''}{b}-(\frac{\,b'}{b})^2-\frac{\,a'}{a}\frac{\,b'}{b}-\frac{\,b'}{b}\frac{\,c'}{c}\right\}\textmd{F}=0.
\end{eqnarray}
Therefore, there are only two field equations containing four unknowns, i.e. the metric coefficients and $\textmd{F}(z)$.
Thus, any set of functions $a(z)$, $b(z)$, $c(z)$ and $\textmd{F}(z)$ satisfying the above two equations would be a solution
of the modified field equations. It is obvious that the solution of these equations could not be found easily. On the other hand, we know that some of the constant curvature solutions in $f(\textmd{R})$ gravity are equal to the solutions in Einstein theory. Hence, in the next section, we will study the simple (but important) case of solutions with constant curvature.
\subsection{Constant curvature solutions}
For the constant curvature solutions, $\textmd{R}=\textmd{R}_0=\textmd{constant}$, we have $\textmd{F}'(\textmd{R}_0)=\textmd{F}''(\textmd{R}_0)=0$. By applying these conditions, equations (10), (11) and (8) respectively are changed to
\begin{eqnarray}
&&2\frac{\,a''}{a}-2\frac{\,b''}{b}-(\frac{\,a'}{a})^2+\frac{\,a'}{a}\frac{\,b'}{b}-\frac{\,a'}{a}\frac{\,c'}{c}+\frac{\,b'}{b}\frac{\,c'}{c}=0,\\
&&2\frac{\,b''}{b}-(\frac{\,b'}{b})^2-\frac{\,a'}{a}\frac{\,b'}{b}-\frac{\,b'}{b}\frac{\,c'}{c}=0,\\
&&2\frac{\,a''}{a}+4\frac{\,b''}{b}-(\frac{\,a'}{a})^2-(\frac{\,b'}{b})^2+2\frac{\,a'}{a}\frac{\,b'}{b}-\frac{\,a'}{a}\frac{\,c'}{c}-2\frac{\,b'}{b}\frac{\,c'}{c}-2\textrm{R}_0c=0.
\end{eqnarray}
It is not an easy task to find the general solutions for these equations. Firstly, by eliminating the variable $a$ from the equations (12) and (13), we lead to the following differential equation
\begin{eqnarray}
2\frac{\,b'''}{b'}-3\frac{\,b''}{b}-\frac{\,c''}{c}+(\frac{\,b'}{b})^2+2(\frac{\,c'}{c})^2-3(\frac{\,b''}{b'}-\frac{1}{2}\frac{\,b'}{b})\frac{\,c'}{c}=0.
\end{eqnarray}
In continuation, calculations show that the following expression
\begin{eqnarray}
\frac{\,b'}{b}=-\frac{4\eta\sqrt{c\,}}{3}\tan(\displaystyle\int\eta\sqrt{c\,}dz+\delta),
\end{eqnarray}
can be a general solution of the equation (15), while $\eta$ and $\delta$ are constants of integration. Moreover, combining this result with the equation $(13)$ yields
\begin{eqnarray}
\frac{\,a'}{a}=\frac{4\eta\sqrt{c\,}}{3}\left\{3\textmd{cosec}2(\displaystyle\int\eta\sqrt{c\,}dz+\delta)-\tan(\displaystyle\int\eta\sqrt{c\,}dz+\delta)\right\}.
\end{eqnarray}
By substituting the equations (16) and (17) into equation (14), after a rather tedious
calculation and simplifying, one obtains$\footnote{Most of the calculations were done using Maple software.}$
\begin{eqnarray}
c'-mc\sqrt{c\,}\sin2(\displaystyle\int\eta\sqrt{c\,}dz+\delta)=0,
\end{eqnarray}
where $m=\frac{\,\textmd{R}_0}{4\eta}+\frac{4\eta}{3}$. By differentiating of this equation, we find that
\begin{eqnarray}
2cc''-3(c')^2-4\eta c\sqrt{m^2c^4-(c')^2c\,}=0.
\end{eqnarray}
Unfortunately only an integral expression as $z=z(c)$ can be obtained from the solution of this differential equation as
\begin{eqnarray}
\int\frac{dc}{\sqrt{8\eta^2(\varepsilon-1)c^3\ln c+(m^2-4\eta^2\varepsilon^2)c^3\,}}=\pm\,z,
\end{eqnarray}
where $\varepsilon$ is an arbitrary constant. This integral equation can be solved exactly only for $\varepsilon=1$. In this case, the existence of the real solutions for $\textmd{R}_0=0$ are impossible and also we must have $\frac{\,\textmd{R}_0}{2}>\frac{4\eta^2}{3}$ or $\frac{\,\textmd{R}_0}{10}<-\frac{4\eta^2}{3}$. However, the solution of integral equation (20) for special case $\varepsilon=1$ becomes
\begin{eqnarray}
c=\frac{4}{(m^2-4\eta^2)z^2}.
\end{eqnarray}
After substituting this expression into equation (16) and integrating, it is found that
\begin{eqnarray}
b=\cos^{\frac{4}{3}}(2\eta\theta),
\end{eqnarray}
where $\theta=\displaystyle\frac{\ln z}{\sqrt{m^2-4\eta^2\,}}$ and we also have taken the constant $\delta$ to be zero without any loss of generality. In order to determine the another metric coefficient, we look at the equation (17). It is difficult to solve this equation. But, we can use the equation (13) which looks simpler. Therefore, after some work, it is given by
\begin{eqnarray}
a=\left(\sin^2(2\eta\theta)\tan(2\eta\theta)\right)^{\frac{2}{3}}.
\end{eqnarray}
By introducing the new variable $\tilde{z}=\displaystyle\frac{\pi}{2}-2\theta$, one can rewrite the metric (7) as follows
\begin{eqnarray}
ds^2=-\cos^2(\eta\tilde{z})\sin^{-\frac{2}{3}}(\eta\tilde{z})dt^2+\sin^{\frac{4}{3}}(\eta\tilde{z})(dx^2+dy^2)+d\tilde{z}^2,
\end{eqnarray}
in which $\eta$ has to be a real odd integer number. This metric has the same general form as Novotn\'{y}-Horsk\'{y} solution with cosmological constant $\Lambda=\frac{4\eta^2}{3}$, [39].
\section{Energy distribution of Novotn\'{y}-Horsk\'{y} solution}
In this section, we calculate the energy distribution of constant curvature solution (24). For doing this, the generalized Landau-Lifshitz energy-momentum complex will be used. We note that this energy-momentum complex is used only for the constant curvature solutions. The calculations show that we are unable to formulate a general expression for the energy-momentum complex which valid for all metrics and theories. The generalized Landau-Lifshitz energy-momentum complex for a general $f(\textmd{R})$ theory is given by, [30]:
\begin{eqnarray}
\tau^{\mu\nu}=\tau^{\mu\nu}_{LL}f'(\textmd{R}_0)+\frac{1}{48\pi\textmd{G}}\left(f'(\textmd{R}_0)\textmd{R}_0-f(\textmd{R}_0)\right)\frac{\partial}{\partial x^\delta}\left(\textrm{g}^{\mu\nu}x^\delta-\textrm{g}^{\mu\delta}x^\nu\right),
\end{eqnarray}
where $\tau^{\mu\nu}_{LL}$ is the Landau-Lifshitz energy-momentum complex evaluated in the framework of general relativity with the following form
\begin{eqnarray}
\tau^{\mu\nu}_{LL}=(-\textrm{g})\left(t^{\mu\nu}_{LL}+\textmd{T}^{\mu\nu}\right),
\end{eqnarray}
and the energy-momentum pseudotensor $t^{\mu\nu}_{LL}$ is defined via the following expression
\begin{eqnarray}
16\pi\textmd{G}\,t^{\mu\nu}_{LL}&=&\left(\textrm{g}^{\mu\alpha}\textrm{g}^{\nu\beta}-\textrm{g}^{\mu\nu}\textrm{g}^{\alpha\beta}\right)\left(2\Gamma^\gamma_{\alpha\beta}\Gamma^\delta_{\gamma\delta}-\Gamma^\gamma_{\alpha\delta}\Gamma^\delta_{\beta\gamma}-
\Gamma^\gamma_{\alpha\gamma}\Gamma^\delta_{\beta\delta}\right)\nonumber\\
&+&\textrm{g}^{\mu\alpha}\textrm{g}^{\beta\gamma}\left(\Gamma^\nu_{\alpha\delta}\Gamma^\delta_{\beta\gamma}+\Gamma^\nu_{\beta\gamma}\Gamma^\delta_{\alpha\delta}-
\Gamma^\nu_{\gamma\delta}\Gamma^\delta_{\alpha\beta}-\Gamma^\nu_{\alpha\beta}\Gamma^\delta_{\gamma\delta}\right)\nonumber\\
&+&\textrm{g}^{\nu\alpha}\textrm{g}^{\beta\gamma}\left(\Gamma^\mu_{\alpha\delta}\Gamma^\delta_{\beta\gamma}+\Gamma^\mu_{\beta\gamma}\Gamma^\delta_{\alpha\delta}
-\Gamma^\mu_{\gamma\delta}\Gamma^\delta_{\alpha\beta}-\Gamma^\mu_{\alpha\beta}\Gamma^\delta_{\gamma\delta}\right)\nonumber\\
&+&\textrm{g}^{\alpha\beta}\textrm{g}^{\gamma\delta}\left(\Gamma^\mu_{\alpha\gamma}\Gamma^\nu_{\beta\delta}-\Gamma^\mu_{\alpha\beta}\Gamma^\nu_{\gamma\delta}\right),
\end{eqnarray}
where $\Gamma^\sigma_{\mu\nu}$ are the usual Christoffel symbols constructed from $\textrm{g}_{\mu\nu}$. The equation (25) is a generalized formula of the Landau-Lifshitz energy-momentum complex which valid for any $f(\textmd{R})$ model with constant curvature solutions. We see that the generalized Landau-Lifshitz energy-momentum complex in $f(\textmd{R})$ theory coincides with the Landau-Lifshitz energy-momentum complex in general relativity only if $f(\textmd{R}_0)=0$ and $f'(\textmd{R}_0)=1$. Next, the 00-component of the relation (25) is given by
\begin{eqnarray}
\tau^{00}=\tau^{00}_{LL}f'(\textmd{R}_0)+\frac{1}{48\pi\textmd{G}}\left(f'(\textmd{R}_0)\textmd{R}_0-f(\textmd{R}_0)\right)(3\textrm{g}^{00}+\frac{\partial \textrm{g}^{00}}{\partial x^i}\,x^i).
\end{eqnarray}
Furthermore, we know that energy of the gravitational field is obtained by the integrated $\tau^{00}$ over the 3-dimensional space integral, [30]:
\begin{eqnarray}
E=\iiint\tau^{00}dx^1dx^2dx^3,
\end{eqnarray}
which is an important quantity of the physical system.\\
In continuation, for determining the 00-component of $t^{\mu\nu}_{LL}$, we need to determine the nonzero Christoffel symbols of metric (24). The calculations show that
\begin{eqnarray}
&\Gamma^0_{03}=\frac{\,2\eta\,}{3}(2\cos^2\textsf{z}-3)\csc\textsf{z},\,\,\,\Gamma^1_{13}=\Gamma^2_{23}=\frac{\,2\eta\,}{3}\cot\textsf{z},\quad&\nonumber\\
&\Gamma^3_{00}=\frac{\,\eta\,}{3}(2\cos^2\textsf{z}-3)\cos\textsf{z}\sin^{-\frac{5}{3}}\textsf{z},\,\,\,\Gamma^3_{11}=\Gamma^3_{22}=-\frac{\,2\eta\,}{3}\cos\textsf{z}\sin^{\frac{1}{3}}\textsf{z},\,\,\,&
\end{eqnarray}
in which $\textsf{z}=\eta\tilde{z}$. In addition, the corresponding Ricci scalar is $\textmd{R}_0=-\frac{16\eta^2}{3}=-4\Lambda$. By substituting the above Christoffel symbols into equation (27), after a rather tedious calculation, one obtains
\begin{eqnarray}
t^{00}_{LL}=-\frac{1}{16\pi\textmd{G}}\,\textmd{g}^{00}\left\{\,\textmd{g}^{00}\Gamma^0_{03}\Gamma^3_{00}+4\textmd{g}^{11}\Gamma^2_{23}\Gamma^3_{11}+\textmd{g}^{33}(\Gamma^0_{03})^2-6\textmd{g}^{33}(\Gamma^1_{13})^2\frac{}{}\right\}.
\end{eqnarray}
After simplifying, this relation takes the simple form
\begin{eqnarray}
t^{00}_{LL}=-\frac{5\Lambda}{24\pi\textmd{G}}\sin^{-\frac{4}{3}}\textsf{z},
\end{eqnarray}
and it yields
\begin{eqnarray}
\tau^{00}_{LL}=-\frac{5\Lambda}{24\pi\textmd{G}}\cos^2\textsf{z}\sin^{\frac{2}{3}}\textsf{z}.
\end{eqnarray}
After inserting this value into equation (28), we finally get
\begin{eqnarray}
\tau^{00}=-\frac{5\Lambda}{24\pi\textmd{G}}\,f'(\textmd{R}_0)\mathbb{Z}_1+\frac{1}{48\pi\textmd{G}}(f'(\textmd{R}_0)\textmd{R}_0-f(\textmd{R}_0))\mathbb{Z}_2, \end{eqnarray}
in which $\mathbb{Z}_1=\cos^2\textsf{z}\sin^{\frac{2}{3}}\textsf{z}$ and $\mathbb{Z}_2=-3\sec^2\textsf{z}\sin^{\frac{2}{3}}\textsf{z}-\textsf{z}\,\frac{d}{d\textsf{z}}(\sec^2\textsf{z}\sin^{\frac{2}{3}}\textsf{z})$. This relation is valid for any $f(\textmd{R})$ theory which has the Novotn\'{y}-Horsk\'{y} metric as a vacuum solution. Below, we will calculate this energy density for some well known $f(\textmd{R})$ models with the constant curvature condition.
\subsection{First model}
At first, we discussed an important $f(\textmd{R})$ model as follows, [26,27]:
\begin{eqnarray}
f(\textmd{R})=\textmd{R}-\frac{\,\,\mu^4\,}{\textmd{R}}-\sigma\textmd{R}^2,
\end{eqnarray}
where $\mu$ and $\sigma$ are real numbers. This model with $\sigma=0$ is the first dark energy model introduced in $f(\textmd{R})$ gravity, called the Carroll-Duvuri-Tordden-Turnner model. It is mentioned here that this $f(\textmd{R})$ model satisfy the constant curvature condition, i.e. $f'(\textmd{R}_0)\textmd{R}_0-2f(\textmd{R}_0)=0$, which implies that $\mu^4=\frac{16\Lambda^2}{3}$. By applying this result, the 00-component of the corresponding generalized Landau-Lifshitz energy-momentum complex is given by
\begin{eqnarray}
\tau^{00}=-\frac{\,1+6\sigma\Lambda\,}{18\pi\textmd{G}}(\mathbb{Z}_2+5\mathbb{Z}_1)\Lambda.
\end{eqnarray}
After integrating, the energy distribution function per unit surface is calculated as
\begin{eqnarray}
E(\textsf{z})=-\frac{(1+6\sigma\Lambda)\sqrt{\Lambda}\,}{9\sqrt{3}\,\pi\textmd{G}}\,\mathbb{Z}\sin^{\frac{2}{3}}\textsf{z},
\end{eqnarray}
where $\mathbb{Z}$ is defined as\vspace{-.8cm}\\
\begin{eqnarray}
&\mathbb{Z}=-\textsf{z}\sec^2\textsf{z}-\frac{1}{8}\tan\textsf{z}(15\sin^2\textsf{z}+1)&\nonumber\\
&\,\,\qquad\qquad+\,\frac{77}{40}\,\textmd{hypergeom}\left(\frac{1}{2},\frac{5}{6},\frac{11}{6};\sin^2\textsf{z}\right)\sin\textsf{z}.&
\end{eqnarray}
\subsection{Second model}
Nojiri and Odintsov [26] suggested a new model of modified gravity which contains the positive and negative powers of curvature as follows
\begin{eqnarray}
f(\textmd{R})=\textmd{R}-(-1)^{n-1}\frac{\alpha}{\,\,\textmd{R}^n}+(-1)^{p-1}\beta\textmd{R}^p,
\end{eqnarray}
where $n$ and $p$ are positive integers and $\alpha$, $\beta$ are any real numbers. They proved that the terms with positive powers of curvature provide the inflationary epoch while the terms with negative powers serves as an alternative for dark energy which is responsible for the cosmic acceleration. This model must satisfy the constant curvature condition, and this condition yields
\begin{eqnarray}
(n+2)\alpha+(p-2)\beta(4\Lambda)^{n+p}=(4\Lambda)^{n+1}.
\end{eqnarray}
For the particular case $p=2$ or $\beta=0$, we get
\begin{eqnarray}
\alpha=\frac{\,\,(4\Lambda)^{n+1}}{n+2},\,\,\,n\not=-2.
\end{eqnarray}
In this case, we have $f(\textmd{R}_0)=\frac{n+1}{n+2}\,\textmd{R}_0$ and $f'(\textmd{R}_0)=\frac{2n+2}{n+2}$. The equation (41) satisfies the constant curvature condition which is necessary for the acceptability of the model (39). After imposing this condition, the energy function takes the following form
\begin{eqnarray}
E(\textsf{z})=-\frac{(n+1)\sqrt{\Lambda}}{6\sqrt{3}\,(n+2)\pi\textmd{G}}\,\mathbb{Z}\sin^{\frac{2}{3}}\textsf{z}.
\end{eqnarray}
\subsection{Third model}
One of the another cosmologically interesting $f(\textmd{R})$ model is given by, [40]:
\begin{eqnarray}
&f(\textmd{R})=\textmd{R}-\varrho\ln(\frac{|\textmd{R}|}{k})+(-1)^{n-1}\zeta\textmd{R}^n,& \end{eqnarray}
where its parameters are related to the cosmological constant. The $f(\textmd{R})$ modified theories with the $\ln\textrm{R}$ term often conduct to a consistent modified gravity which may pass the Solar System tests, [40,41]. The constant scalar curvature condition gives
\begin{eqnarray}
&\varrho-4\Lambda-2\varrho\ln\left(\frac{4\Lambda}{k}\right)+(n-2)\zeta(4\Lambda)^n=0.&
\end{eqnarray}
For the case $n=2$ or $\zeta=0$, this condition reduces to
\begin{eqnarray}
\varrho=\frac{4\Lambda}{1-2\ln(\frac{4\Lambda}{k})},
\end{eqnarray}
which satisfies the constant curvature condition necessary for acceptability of the
model (43). By inserting the value $f'(\textmd{R}_0)=\frac{2\ln(\frac{4\Lambda}{k})}{2\ln(\frac{4\Lambda}{k})-1}$ into equation (34), we obtain
\begin{eqnarray}
\tau^{00}=-\frac{\,\mathbb{Z}_2+\ln(\frac{4\Lambda}{k})(\mathbb{Z}_2+5\mathbb{Z}_1)\,}{12\pi\textmd{G}\left(2\ln(\frac{4\Lambda}{k})-1\right)}\Lambda.
\end{eqnarray}
Finally, the energy distribution function for the model (43) is obtained as follows
\begin{eqnarray}
E(\textsf{z})=-\displaystyle\frac{\,\mathbb{Z}\ln(\frac{4\Lambda}{k})-\textsf{z}\sec^2\textsf{z}-2\tan\textsf{z}
+\frac{4}{5}\,\textmd{hypergeom}\left(\frac{1}{2},\frac{5}{6},\frac{11}{6};\sin^2\textsf{z}\right)\sin\textsf{z}\,}{6\sqrt{3}\,\pi\textmd{G}\left(2\ln(\frac{4\Lambda}{k})-1\right)}\sin^{\frac{2}{3}}\textsf{z}\sqrt{\Lambda}.
\end{eqnarray}
\newpage
\section{Conclusion}
In this work, the exact solutions of a general plane symmetric spacetime have been investigated in the framework of metric $f(\textmd{R})$ gravity. Firstly, it is found that the vacuum solutions with constant curvature are exactly similar to the Novotn\'{y}-Horsk\'{y} solution with a parameter which is identified as the cosmological constant. For this solution, the energy distribution functions have been calculated for some important $f(\textmd{R})$ models by using the generalized Landau-Lifshitz energy-momentum complex. It was also found that the constant curvature condition is satisfied for these models.\vspace{.2cm}
\begin{eqnarray*}
{\bf{References}}
\end{eqnarray*}
1. Einstein, A.: Sitzungsber. Preus. Akad. Wiss. Berlin. {\bf47}, 778 (1915).\\
2. Papapetrou, A.: Proc. R. Irish Acad. A {\bf52}, 11 (1948).\\
3. Bergmann, P.G., Thomson, R.: Phys. Rev. {\bf89}, 400 (1953).\\
4. Goldberg, J.N.: Phys. Rev. {\bf111}, 315 (1958).\\
5. Weinberg, S.: Gravitation and Cosmology (Wiley, New York, 1972).\\
6. M{\o}ller, C.: Ann. Phys. {\bf4}, 347 (1958).\\
7. Landau, L.D., Lifshitz, E.M.: The Classical Theory of Fields (Addison-Wesley Press, New York, 1962).\\
8. Chang, C.C., Nester, J.M., Chen, C.: Phys. Rev. Lett. {\bf83}, 1897 (1999).\\
9. Hawking, S.W., Horowitz, G.T.: Class. Quantum Grav. {\bf13}, 1487 (1996).\\
10. Liu, C.-C.M., Yau, S.-T.: Phys. Rev. Lett. {\bf90}, 231102 (2003).\\
11. Cooperstock, F.I., Sarracino, R.S.: J. Phys. A: Math. Gen. {\bf11}, 877 (1978).\\
12. Aguirregabiria, J.M., Chamorro, A., Virbhadra, K.S.: Gen. Relativ. Grav. {\bf28}, 1393 (1996).\\
13. Riess, A.G., et al.: Astrophys. J. {\bf607}, 665 (2004).\\
14. Knop, R.A., et al.: Astrophys. J. {\bf598}, 102 (2003).\\
15. Spergel, D.N., et al.: Astrophys. J. Suppl. Ser. {\bf148}, 175 (2003).\\
16. Spergel, D.N., et al.: Astrophys. J. Suppl. Ser. {\bf170}, 377 (2007); Komatsu, E., et al.: Astrophys. J. Suppl. Ser. {\bf180}, 330 (2009).\\
17. Bennett, C.L., et al.: Astrophys. J. Suppl. Ser. {\bf148}, 1 (2003).\\
18. Carroll, S.M.: Living Rev. Relativ. {\bf4}, 1 (2001).\\
19. Copeland, E.J., Sami, M., Tsujikawa, S.: Int. J. Mod. Phys. D {\bf15}, 1753 (2006).\\
20. Padmanabhan, T.: Phys. Rep. {\bf380}, 235 (2003).\\
21. Schmidt, H.J.: Int. J. Geom. Meth. Mod. Phys. {\bf4}, 209 (2007).\\
22. Stelle, K.S.: Phys. Rev. D {\bf16}, 953 (1977).\\
23. Birrell, N.D., Davies, P.C.W.: Quantum Fields in Curved Spacetime (Cambridge University Press, Cambridge, 1982).\\
24. Elizalde, E., Nojiri, S., Odintsov, S.D.: Phys. Rev. D {\bf70}, 043539 (2004).\\
25. Buchdahl, H.A.: Mon. Not. R. Astron. Soc. {\bf150}, 1 (1970).\\
26. Nojiri, S., Odintsov, S.D.: Phys. Rev. D {\bf68}, 123512 (2003).\\
27. Nojiri, S., Odintsov, S.D.: Int. J. Geom. Meth. Mod. Phys. {\bf4}, 115 (2007).\\
28. Nojiri, S., Odintsov, S.D.: Phys. Rev. D {\bf77}, 026007 (2008).\\
29. Faraoni, V.: Phys. Rev. D {\bf74}, 023529 (2006).\\
30. Multam\"{a}ki, T., Putaja, A., Vilja, I., Vagenas, E.C.: Class. Quantum Grav. {\bf25}, 075017 (2008).\\
31. Sharif, M., Shamir, M.F.: Gen. Relativ. Grav. {\bf42}, 1557 (2010).\\
32. Capozziello, S., Francaviglia, M.: Gen. Relativ. Grav. {\bf40}, 357 (2008).\\
33. Sotiriou, T.P., Faraoni, V.: Rev. Mod. Phys. {\bf82}, 451 (2010).\\
34. Olmo, G.J.: Int. J. Mod. Phys. D {\bf20}, 413 (2011).\\
35. Novotn\'{y}, J., Horsk\'{y}, J.: Czech. J. Phys. B {\bf24}, 718 (1974).\\
36. Sharif, M., Shamir, M.F.: Mod. Phys. Lett. A {\bf25}, 1281 (2010).\\
37. Yavari, M.: Astrophys. Space Sci. {\bf348}, 293 (2013).\\
38. Multam\"{a}ki, T., Vilja, I.: Phys. Rev. D {\bf74}, 064022 (2006).\\
39. Stephani, H., Kramer, D., MacCallum, M.A.H., Hoenselaers, C., Herlt, E.: Exact Solutions of Einstein's Field Equations (Cambridge University Press, Cambridge, 2003).\\
40. Nojiri, S., Odintsov, S.D.: Gen. Relativ. Grav. {\bf36}, 1765 (2004).\\
41. Nojiri, S., Odintsov, S.D.: Mod. Phys. Lett. A {\bf19}, 627 (2004).
\end{document}